\documentclass{article}
\usepackage{romp}

\title{ Understanding Partial $\mathcal{PT}$ symmetry as Weighted Composition Conjugation in Reproducing Kernel Hilbert Space :An application to Non-hermitian Bose-Hubbard Type Hamiltonian in Fock space  }
\author{ Arindam Chakraborty \\ Department of Physics, Heritage Institute of Technology, Kolkata-700107, India \\ arindam.chakraborty@heritageit.edu:  \\[2ex]
           }

\begin{document}

\maketitle
\begin{abstract}
     A new kind of symmetry behaviour introduced as partial ${\mathcal {PT}}$-symmetry(henceforth $\partial_{\mathcal{PT}}$) is investigated in a typical Fock space setting understood as  a Reproducing Kernel Hilbert Space (RKHS). The same kind of symmetry is understood for a non-hermitian Bose-Hubbard type Hamiltonian involving two boson operators as well as its eigenstates. The phenomenon of symmetry breaking has also been considered.
\end{abstract}


\section{Introduction}
Non-hermitian Hamiltonian with real eihenvalues in the context of  ${\mathcal {PT}}$ symmetry has become an interesting area of investigation for last couple of decades \cite{bender98, bender99, bender02, brody16, mosta02, brody14,moise11,baga15}. The present article stems from a recent study of partial $\mathcal{PT}$ symmetry by Beygi et. al. \cite{beygi15} where a variable specific action of symmetry operator is understood considering a model of an N-coupled harmonic oscillator Hamiltonian with purely imaginary coupling terms. It has also been observed that the reality and partial reality of the spectrum have direct correspondences with the classical trajectories.The present formulation attempts to explore the possibility of partial $\mathcal{PT}$ symmetry in a Bose-Hubberd hamiltonian operator\cite{graefe08} as well as in its eigenstates in a typical Fock space environment. The relevant Fock space \cite{zhu12} has been viewed as a Reproducing Kernel Hilbert Space \cite{paulsen16,hai18, hai16, hai18c, gar05, gar07}and the symetry operators are understood as Weighted Composition Conjugation \cite{hai18, hai16, hai18c} acting on it.

We begin with the following definition of Fock space involving functions of $n$ complex variable.
\begin{definition}{Definition}
	A Fock (or Segal-Bargmann) space $(\mathcal{F}^2({C}^n))$ is a separable complex Hilbert space of entire functions (of the complex variables $\{\zeta_j : j=1\dots n\}$ ) equipped with an inner-product 
	\begin{eqnarray}
		\langle \psi, \phi\rangle=\prod_{j=1}^n\int_{W(\zeta_j)} \psi(\zeta_j : j=1\dots n)\overline{\phi(\zeta_j : j=1\dots n)}\nonumber\\
		\;\; {\rm{with}} \prod_{j=1}^n\int_{W(\zeta_j)}\equiv \int\int d{W(\zeta_1)}\dots d{W(\zeta_n)}.
	\end{eqnarray}
 Here, ${d}{W(u)}=\frac{1}{\pi}e^{-\vert u\vert^2}{d}({\rm{Re}}(u)) d({\rm{Im}}(u))$ represents the relevant Gaussian measure relative to the complex variable $u$.  
\end{definition}
In a Fock space of one complex variable  ${\mathcal {PT}}$ symmetry is often understood as a consequence of the more general notion of \textbf{weighted composition conjugation} \cite{hai18}  defined as follows.

\begin{definition}{Definition}
	Let, $\zeta$ is a complex variable and $\{\vartheta, \eta, \upsilon\}$ are complex numbers satisfying the set of necessary and sufficient conditions : $\vert\vartheta\vert=1,\bar{\vartheta}\eta+\bar{\eta}=0$ and $\vert\upsilon\vert^2e^{\vert\eta\vert^2}=1$.The weighted composition conjugation is defined as	

\begin{eqnarray}
	\mathcal{C}_{(\vartheta, \eta, \upsilon)}\psi(\zeta)=\upsilon e^{\eta\zeta}\overline{\psi(\overline{\vartheta\zeta+\eta})}.
\end{eqnarray}
\end{definition}
The anti-linear operator $\mathcal{C}_{(\vartheta, \eta, \upsilon)}$ is a \textbf{conjugation} since it is involutive and isometric. The action of the operator $\mathcal{PT}$ is equivalent to the choice : $\vartheta=-1=-\upsilon,\eta=0$ which results to the following equation
\begin{eqnarray}
	\mathcal{C}_{(\vartheta, 0, 1)}\vert_{\vartheta=-1}\psi(\zeta)=\overline{\psi(\overline{-\zeta})}.
\end{eqnarray}
Similarly, the action of $\mathcal{T}$ is indicative of the choice : $\vartheta=1, \eta=0, \upsilon=1$ giving
\begin{eqnarray}
	\mathcal{C}_{(\vartheta, 0, 1)}\vert_{\vartheta=1}\psi(\zeta)=\overline{\psi(\overline{\zeta})}.
\end{eqnarray}
If $\psi$ is a function of several complex variables $\{\zeta_j : j=1\dots n\}$ one can define an operator $\mathcal{C}_{(\vartheta_j,\eta_j, \upsilon_j : j=1\dots n )}$ with the action
\begin{eqnarray}
	\mathcal{C}_{(\vartheta_j,\eta_j= 0,\upsilon_j= 1 : j=1\dots n)}\psi(\zeta_1,\dots,\zeta_j,\dots,\zeta_n)=\overline{\psi(\overline{\vartheta_1\zeta_1},\dots,\overline{\vartheta_j\zeta_j},\dots, \overline{\vartheta_n\zeta_n})}.
\end{eqnarray} 
Let us introduce an operator $\mathcal{C}^{(j)}_n=\mathcal{C}_{(\vartheta_j,\eta_j= 0,\upsilon_j= 1 ; j=1\dots n)}\vert_{\vartheta_1=1,\dots,\vartheta_j=-1,\dots,\vartheta_n=1}$ as $j$-th partial $\mathcal{PT}$ symmetry ($\partial_{\mathcal{PT}}$) operator through the following action

\begin{eqnarray}
	\mathcal{C}^{(j)}_n\psi(\zeta_1,\dots,\zeta_j,\dots,\zeta_n)=\overline{\psi(\bar{\zeta_1},\dots,\overline{-\zeta_j},\dots,\bar{\zeta_n})}.
\end{eqnarray}
and a global $\mathcal{PT}$ symmetry operator $\mathcal{C}_n$ through the action
\begin{eqnarray}
	\mathcal{C}_n\psi(\zeta_1,\dots,\zeta_j,\dots,\zeta_n)=\overline{\psi(\overline{-\zeta_1},\dots,\overline{-\zeta_j},\dots,\overline{-\zeta_n})}.
\end{eqnarray}
For our present purpose we shall only consider the operators $\mathcal{C}_2$ and $\{\mathcal{C}_2^{(j)} : j=1,2\}$. Now, global and partial  $\mathcal{PT}$ symmetries of any
function $\psi(\zeta_1, \zeta_2)$ are understood through the following equations
\begin{eqnarray}
	\mathcal{C}_2\psi(\zeta_1, \zeta_2)=\psi(\zeta_1, \zeta_2)\:\:{\rm and}\:\:\mathcal{C}_2^{(j)}\psi(\zeta_1, \zeta_2)=\psi(\zeta_1, \zeta_2)\:\:\forall\:\: j=1, 2
\end{eqnarray}
respectively.
\section{The model Hamiltonian and $\partial_{\mathcal{PT}}$ symmetry in Fock space}
In the present discussion, following \cite{graefe08} a Bose-Hubbard type Hamiltonian has been considered. Such a Hamiltonian has been invoked as a two mode version for a second quantized many particle system showing Bose-Einstein Condensation (BEC) in a double well potential at low temperature. The said Hamiltonian becomes non-hermitian if one of the interaction terms present in it is taken as purely imaginary.The model Hamiltonian under consideration is given by
\begin{eqnarray}
	H=\epsilon_0(a_1^{\dagger}a_1-a_2^{\dagger}a_2)+\epsilon(a_1^{\dagger}a_2+a_2^{\dagger}a_1)+\alpha(a_1^{\dagger}a_1-a_2^{\dagger}a_2)^2.
\end{eqnarray}
Here $\epsilon_0$ represents the on site energy difference, $\epsilon$ is the single particle tunneling and $\alpha$ stands for the interaction strength. $\{a_j, a_j^{\dagger} : j=1, 2\}$ are boson operators satisfying the condition $[a_j, a^\dagger_k] - \delta_{jk}=[a_{j}, a_{k}]=[a^{\dagger}_{j}, a^{\dagger}_{k}]=0$. The Hamiltonian commutes with the number operator $N=a_1^{\dagger}a_1+a_2^{\dagger}a_2$ indicating particle conservation. For the time being we consider $\epsilon_0=1, \epsilon=i\gamma$ and $\gamma$ and $\alpha$ to be real.

In order to understand $\partial_{\mathcal{PT}}$ symmetry in Fock space we rewrite the Hamiltonian using Bargmann-Fock correspondence : $a_j^{\dagger}=\zeta_j, a_j=\partial_{\zeta_j}$ as follows
\begin{eqnarray}
	H=(\zeta_1\partial_{\zeta_1}-\zeta_2\partial_{\zeta_2})+i\gamma(\zeta_1\partial_{\zeta_2}+\zeta_2\partial_{\zeta_1})+\alpha(\zeta_1\partial_{\zeta_1}-\zeta_2\partial_{\zeta_2})^2.
\end{eqnarray}

Following \cite{hai18, hai16, hai18c} we shall  demonstrate the actions of weighted composition conjugations $\mathcal{C}_2$ and $\mathcal{C}_2^{(j)}$ on $H$ via the notion of Reproducing Kernel Hilbert Space (RKHS).
\begin{definition}{Definition}
	A function of the form $K^{[m_j]}_{\{\zeta_j\}}(u_j:j=1\dots n)=\prod_{j=1}^nu_j^{m_j}e^{u_j\overline{\zeta_j}}$($m_j\in{N}$ and $\zeta_j,u_j\in {C}\:\:\forall\:\:j=1\dots n$) is called a kernel function (or a \textbf{reproducing kernel}) which satisfies the condition 
	\begin{eqnarray}
		\psi^{(m_1,m_2,\dots, m_n)}(\zeta_1, \zeta_2,\dots,\zeta_n)=\langle\psi, K^{[m_j]}_{\{\zeta_1\}}\rangle\nonumber\\
		=\int_{W(u_1)}\int_{W(u_2)}\dots \int_{W(u_n)}\psi(u_1, u_2,\dots,u_n)\overline{K^{[m_j]}_{\{\zeta_1\}}}. 
	\end{eqnarray}
\end{definition} 
Considering the case with two complex variables the following proposition is immediate.

\begin{proposition}{Proposition}
	$H^{\star}\neq H$ where $H^{\star}$ is defined as $\langle H\psi(u_1,u_2), K^{[m_1, m_2]}_{\zeta_1, \zeta_2}\rangle=\langle\psi(u_1,u_2), H^{\star}K^{[m_1, m_2]}_{\zeta_1, \zeta_2}\rangle$. 
\end{proposition}

{\it Proof} : 	We shall first show that $\langle u_1\partial_{u_1}\psi(u_1,u_2), K^{[m_1, m_2]}_{\zeta_1, \zeta_2}\rangle=\langle\psi(u_1,u_2), u_1\partial_{u_1}K^{[m_1, m_2]}_{\zeta_1, \zeta_2}\rangle $.
\begin{eqnarray}
		\langle  u_1\partial_{u_1}\psi(u_1,u_2), K^{[m_1, m_2]}_{\zeta_1, \zeta_2}\rangle=\langle u_1\partial_{u_1}\psi{(u_1, u_2)},u_1^{m_1}u_2^{m_2}e^{u_1\overline{\zeta_1}+u_2\overline{\zeta_2}}\rangle\nonumber\\
		=\int_{W(u_1)}\int_{W(u_2)}u_1\partial_{u_1}\psi(u_1, u_2)\overline{u_1}^{m_1}\overline{u_2}^{m_2}e^{\overline{u_1}{\zeta_1}+\overline{u_2}{\zeta_2}}\nonumber\\
		=\int_{W(u_1)}\int_{W(u_2)}u_1[\int_{W(v_1)}\int_{W(v_2)}\overline{v_1}\psi(v_1, v_2)e^{\overline{v_1}{u_1}+\overline{v_2}{u_2}}]\overline{u_1}^{m_1}\overline{u_2}^{m_2}e^{\overline{u_1}{\zeta_1}+\overline{u_2}{\zeta_2}}\nonumber\\
		=\int_{W(v_1)}\int_{W(v_2)}\overline{v_1}\psi(v_1, v_2)[\int_{W(u_1)}\int_{W(u_2)}u_1\overline{u_1}^{m_1}\overline{u_2}^{m_2}e^{\overline{u_1}{\zeta_1}+\overline{u_2}{\zeta_2}}e^{\overline{v_1}{u_1}+\overline{v_2}{u_2}}]\nonumber\\
		=\int_{W(v_1)}\int_{W(v_2)}\overline{v_1}\psi(v_1, v_2)\langle u_1e^{\overline{v_1}{u_1}+\overline{v_2}{u_2}}, {u_1}^{m_1}{u_2}^{m_2}e^{{u_1}\overline{\zeta_1}+{u_2}\overline{\zeta_2}}\rangle\nonumber\\
		= \int_{W(v_1)}\int_{W(v_2)}\overline{v_1}\psi(v_1, v_2)\partial_{\zeta_1}^{m_1}\partial_{\zeta_2}^{m_2}(\zeta_1e^{\overline{v_1}{\zeta_1}+\overline{v_2}{\zeta_2}})\nonumber\\
		= \int_{W(v_1)}\int_{W(v_2)}\overline{v_1}\psi(v_1, v_2)\partial_{\zeta_1}^{m_1}\partial_{\zeta_2}^{m_2}\partial_{\overline{v_1}}(e^{\overline{v_1}{\zeta_1}+\overline{v_2}{\zeta_2}})\nonumber\\
		=\int_{W(v_1)}\int_{W(v_2)}\psi(v_1, v_2)\overline{v_1}\partial_{\overline{v_1}}\partial_{\zeta_1}^{m_1}\partial_{\zeta_2}^{m_2}(e^{\overline{v_1}{\zeta_1}+\overline{v_2}{\zeta_2}})\nonumber\\
=\langle\psi(v_1, v_2), v_1\partial_{v_1}K^{[m_1, m_2]}_{\zeta_1, \zeta_2}(v_1, v_2)\rangle.
\end{eqnarray}
	
	An identical argument holds for $ u_2\partial_{u_2}$ and similar calculations justify forms of adjoints for the operators $i u_j\partial_{u_k}$ for $j\neq k$. For example, it can be shown that
	\begin{eqnarray}
		\langle iu_1\partial_{u_2}\psi(u_1, u_2), u_1^{m_1}u_2^{m_2}e^{u_1\overline{\zeta_1}+u_2\overline{\zeta_2}}\rangle\nonumber\\
		=\langle\psi(v_1, v_2), -iv_2\partial_{v_1}v_1^{m_1}v_2^{m_2}e^{v_1\overline{\zeta_1}+v_2\overline{\zeta_2}}\rangle.
	\end{eqnarray}
	Using these results in the expression of Hamiltonian the proposition is verified.\rule{5pt}{5pt}

\begin{proposition}{Proposition}
	$H$ is $\mathcal{C}_2$ self-adjoint i. e.; $\mathcal{C}_2H^{\star}\mathcal{C}_2=H$ 		
\end{proposition}
{\it Proof} : We shall first show the case with $u_1\partial_{u_1}$. Considering the conjugation operator $\mathcal{C}_{2}$ we find
\begin{eqnarray}
	\mathcal{C}_{2}(u_1\partial_{u_1})^{\star}\mathcal{C}_{2}K^{[m_1, m_2]}_{\zeta_1, \zeta_2}(u_1, u_2)\nonumber\\
	=\mathcal{C}_{2}u_1\partial_{u_1}\mathcal{C}_{2}u_1^{m_1}u_2^{m_2}e^{u_1\overline{\zeta_1}+u_2\overline{\zeta_2}}\nonumber\\
	=\mathcal{C}_{2}u_1\partial_{u_1}\overline{\overline{(-u_1)}^{m_1}}\overline{\overline{(-u_2)}^{m_2}}\overline{e^{\overline{-u_1}\overline{\zeta_1}+\overline{-u_2}\overline{\zeta_2}}}\nonumber\\
	=\mathcal{C}_{2}u_1\partial_{u_1}(-1)^{m_1+m_2}u_1^{m_1}u_2^{m_2}e^{-u_1{\zeta_1}-u_2{\zeta_2}}\nonumber\\
	=\mathcal{C}_{2}u_1(-1)^{m_1+m_2}[u_1^{m_1}u_2^{m_2}(-{\zeta_1})e^{-u_1{\zeta_1}-u_2{\zeta_2}}+m_1u_1^{m_1-1}u_2^{m_2}e^{-u_1{\zeta_1}-u_2{\zeta_2}}]\nonumber\\
	=-u_1(-1)^{m_1+m_2}[(-1)^{m_1+m_2}u_1^{m_1}u_2^{m_2}(-\overline{\zeta_1})\nonumber\\
	+m_1(-1)^{m_1+m_2-1}u_1^{m_1-1}u_2^{m_2}]e^{u_1\overline{\zeta_1}+u_2\overline{\zeta_2}}\nonumber\\
	=u_1\partial_{u_1}K^{[m_1, m_2]}_{\zeta_1, \zeta_2}(u_1, u_2).
\end{eqnarray}

Similarly one can prove $\mathcal{C}_{2}(iu_1\partial_{u_2})^{\star}\mathcal{C}_{2}K^{[m_1, m_2]}_{\zeta_1, \zeta_2}(u_1, u_2)=(iu_2\partial_{u_1})K^{[m_1, m_2]}_{\zeta_1, \zeta_2}(u_1, u_2)$ and $\mathcal{C}_{2}(iu_2\partial_{u_1})^{\star}\mathcal{C}_{2}K^{[m_1, m_2]}_{\zeta_1, \zeta_2}(u_1, u_2)=(iu_1\partial_{u_2})K^{[m_1, m_2]}_{\zeta_1, \zeta_2}(u_1, u_2)$. Using these result in the expression of the Hamiltonian the above proposition is verified. \rule{5pt}{5pt}

\begin{proposition}{Proposition}
$H$ has $\partial_{\mathcal{PT}}$ symmetry i. e.; $\mathcal{C}_2^{(j)}H\mathcal{C}_2^{(j)}=H$, for $j=1, 2$ but it lacks global $\mathcal{PT}$ symmetry i. e.; 
$\mathcal{C}_2H\mathcal{C}_2\neq H$.  
\end{proposition}	
{\it Proof} :	
We shall only verify the that $\mathcal{C}_2^{(1)}u_1\partial_{u_1}\mathcal{C}_2^{(j)}K^{[m_1, m_2]}_{\zeta_1, \zeta_2}(u_1, u_2)=u_1\partial_{u_1}K^{[m_1, m_2]}_{\zeta_1, \zeta_2}(u_1, u_2)$ through the following steps
\begin{eqnarray}	
	\mathcal{C}_2^{(1)}u_1\partial_{u_1}\mathcal{C}_2^{(1)}K^{[m_1, m_2]}_{\zeta_1, \zeta_2}(u_1, u_2)\nonumber\\
	=\mathcal{C}_2^{(1)}u_1\partial_{u_1}[(-u_1)^{m_1}(u_2)^{m_2}e^{-u_1\zeta_1+u_2\zeta_2}]\nonumber\\
	=\mathcal{C}_2^{(1)}u_1(-1)^{m_1}[m_1u_1^{m_1-1}u_2^{m_2}+(-\zeta_1)u_1^{m_1}u_2^{m_2}]e^{-u_1\zeta_1+u_2\zeta_2}\nonumber\\
	=-u_1(-1)^{m_1}[m_1(-u_1)^{m_1-1}u_2^{m_2}+\overline{(-\zeta_1)}(-u_1)^{m_1}u_2^{m_2}]e^{u_1\overline{\zeta_1}+u_2\overline{\zeta_2}}\nonumber\\
	=u_1\partial_{u_1}K^{[m_1, m_2]}_{\zeta_1, \zeta_2}(u_1, u_2).
\end{eqnarray}
Similarly,
\begin{eqnarray}
	\mathcal{C}_2^{(1)}iu_1\partial_{u_2}\mathcal{C}_2^{(1)}K^{[m_1, m_2]}_{\zeta_1, \zeta_2}(u_1, u_2)\nonumber\\
	=\mathcal{C}_2^{(1)}iu_1\partial_{u_2}[(-u_1)^{m_1}(u_2)^{m_2}e^{-u_1\zeta_1+u_2\zeta_2}]\nonumber\\
	=\mathcal{C}_2^{(1)}iu_1(-1)^{m_1}[(u_1)^{m_1}m_2(u_2)^{m_2-1}+(u_1)^{m_1}(u_2)^{m_2}(\zeta_2)]e^{-u_1\zeta_1+u_2\zeta_2}\nonumber\\
	=iu_1\partial_{u_2}K^{[m_1, m_2]}_{\zeta_1, \zeta_2}(u_1, u_2).
\end{eqnarray}
Using these and similar results for the expression of the Hamiltonian the proposition can be verified.\rule{5pt}{5pt}

\section{$\partial_{\mathcal{PT}}$ symmetry of the eigenstates of the Hamiltonian $H$}
First we shall show that the reality of the eigenvalues is directly related to the $\partial_{\mathcal{PT}}$ symmetry of the eigen states of the Hamiltonian. It is readily observed that the present Hamiltonian leaves the homogeneous polynomial space of two indeterminates $(\zeta_1, \zeta_2)$ invariant. Considering such a space of degree of homogeneity $m$ and polynomial bases $\{f_{k} = \zeta_1^{m-k} \zeta_2^{k}
: k = 0 \dots m\}$ the operator $H$ has the following tridiagonal representation
\begin{eqnarray}\label{tri1}
	H 
	= 
	\left( \begin{array}{cccccccc}
		\beta_m & i\gamma & 0 & 0 & 0& \dots  & 0& 0 \\
		im\gamma & \beta_{m-2} & 2i\gamma & 0 & 0 &\dots &0 & 0 \\
		0& i\gamma(m-1) & \beta_{m-4} & 3i\gamma & 0 & \dots &0 & 0 \\
		\vdots & \vdots & \vdots & \vdots & \vdots & \dots & \beta_{-(m-2)} & im\gamma \\
		0 & 0& 0& 0& 0& \dots & i\gamma & \beta_{-m}
	\end{array} \right).  
\end{eqnarray}
Here, $\beta_{\mu}=\mu+\alpha{\mu}^2$.
The the eigenvalues of such a matrix can be found out with the help of the following theorem \cite{sandry13}.
\begin{theorem}{Theorem}
	Given a tri-diagonal matrix
	\begin{equation}\label{tri2}
		{\mathcal M} = \left(
		\begin{array}{cccccccc}
			b_0 & d_0 & 0 & 0 & 0 & \dots & 0 & 0 \\
			c_0 & b_1 & d_1 & 0 & 0 & \dots & 0 & 0 \\
			0 & c_1 & b_2 & d_2 & 0 & \dots & 0 & 0 \\
			\vdots & \vdots & \vdots & \vdots & \vdots & \dots & b_{l-2} & d_{l-2} \\
			0 & 0 & 0 & 0 & 0 & \dots & c_{l-2} & b_{l-1}
		\end{array}\right)
	\end{equation}
with $d_i \neq 0 \:\:\forall\:\: i$, let us consider a polynomial $P_n (\lambda)$
that follows the well-known three term recursion relation \cite{dunkl14, ismail05} 
\begin{equation}\label{rec1}
	P_{n+1}(\lambda) = \frac{1}{d_n}[(\lambda - b_n )P_n (x) - c_{n-1} P_{n-1} (\lambda)].    
\end{equation}
If $P_{-1}(\lambda) = 0$ and $P_0(\lambda) = 1$ the eigenvalues are given by the zeros
of the polynomial $P_l(\lambda)$ and eigenvector corresponding to $j$-th eigenvalue
$\lambda_j$ is given by the vector
\begin{equation}
	\left(
	\begin{array}{c}
		P_0 (\lambda_j ) \\
		P_1 (\lambda_j )\\
		\vdots \\
		P_{l-2} (\lambda_j)\\
		P_{l-1} (\lambda_j )      
	\end{array}\right).
\end{equation}
\end{theorem}
{\it Proof} :
	Let $
	\left(
	\begin{array}{c}
		v_0  \\
		v_1 \\
		\vdots \\
		v_{l-2} \\
		v_{l-1}       
	\end{array}\right).
	$be the eigen vector corresponding to the eigenvalue $\lambda$. Then the eigenvalue equation gives us
	\begin{eqnarray}
		b_0v_0+d_0v_1=\lambda v_0\nonumber\\
		c_0v_0+b_1v_1+d_1v_2=\lambda v_1\nonumber\\
		\vdots\nonumber\\
		c_{l-2}v_{l-2}+b_{l-1}v_{l-1}=\lambda v_{l-1}
	\end{eqnarray}
	Since $P_0(\lambda)=1$, one can write $v_0=P_0(\lambda)v_0$. Now in view of the recurrence relation (equation-\ref{rec1}) 
\begin{eqnarray}
	v_1=\frac{1}{d_0}(\lambda-b_0)P_0(\lambda)v_0=P_1(\lambda)
\end{eqnarray} 
Continuing the substitution recursively we get 
\begin{equation}\label{rec2}
	v_n=P_n(\lambda)v_0 : 0\leq n <l
\end{equation}
Substituting $n=l-1$ in equation-\ref{rec2} and using the three term relation (equation-\ref{rec1}) we get
\begin{equation}
	P_l(\lambda)v_0=0
\end{equation}
giving the characteristic equation $P_l(\lambda)=0$.\rule{5pt}{5pt}

Now going back to the matrix in equation-\ref{tri1}
and comparing the matrices in  equation-\ref{tri1} and equation-\ref{tri2} we get $\{b_0,\dots, b_{l-1}\}=\{\beta_m,\dots,\beta_{-m}\}$, $\{c_0,\dots, c_{l-2}\}=\{im\gamma,\dots,i\gamma\}$ and $\{d_0,\dots, d_{l-2}\}=\{i\gamma,\dots,im\gamma\}$, one can determine the eigenvectors and eigenvalues of the matrix using the above algorithm.
\subsection{\textbf{Reality of the eigenvalues and $\partial_{\mathcal{PT}}$ symmetry of the eigenstates}}
Without calculating the eigenvalues explicitly, an immediate inference regarding the symmetry behaviour of the eigenfunctions is possible in view of the reality of the eigenvalues. We shall consider the following two cases.

\vspace{0.5cm}

\textbf{Case-I} :  $\vert m\vert\in\{2s : s\in {Z^+}\}$
Let us begin from the value  $\vert m\vert=2$. The matrix  of $H$ is a $3\times 3$ matrix. Correspondingly the matrix ${\mathcal M}$ has the diagonal elements $b_0=2+4\alpha, b_1=0, b_2=-2+4\alpha$. This implies $l=3$. When $\vert m\vert=2s$ for some fixed $s$, $l=2s+1$ and the matrix ${\mathcal M}$ becomes a $(2s+1)\times (2s+1)$ matrix. As a consequence a polynomial equation $P_l(x)=0$ implies $P_{2s+1}(x)=0$ and roots of this polynomial equation gives us the eigenvalues. The eigenvector corresponding to an eigenvalue $\lambda_0$ would be given by the vector
\begin{eqnarray}
	(P_0 (\lambda_0 )\:\:\:P_1 (\lambda_0 )\dots P_{2s} (\lambda_0 )).
\end{eqnarray}
The first term is equal to one for all values of $\lambda_0$. Now for real $\lambda_0$ it is easy to verify that starting from the purely imaginary second term the subsequent terms are alternatively real and purely imaginary leading to the following equivalence
\begin{eqnarray}
	(P_0 (\lambda_0 )\:\:\:P_1 (\lambda_0 )\dots P_{2s} (\lambda_0 ))=(A_0 (\lambda_0 )\:\:\:iA_1 (\lambda_0 )\dots A_{2s} (\lambda_0 )).
\end{eqnarray}
Here, $\{A_l : l=0\dots 2s\}$ are real functions of $\lambda_0$ with $A_0=1$. In the Fock space setting the eigenfunction in $({\zeta_1, \zeta_2})$ can be given by
\begin{eqnarray}
	\psi^{(I)}_{2s}(\zeta_1, \zeta_2)=A_0\zeta_1^{2s}+iA_1\zeta_1^{2s-1}\zeta_2+A_2\zeta_1^{2s-2}\zeta_2^{2}+\dots +A_{2s}\zeta_2^{2s}.
\end{eqnarray}
Now the action of partial parity operator $\{\mathcal{C}_2^{(j)} : j=1, 2 \}$ on $\psi^{(I)}_{2s}(\zeta_1, \zeta_2)$ can be understood through the following equation
\begin{eqnarray}
	\mathcal{C}_2^{(1)}\psi^{(I)}_{2s}(\zeta_1, \zeta_2)\nonumber\\
	=\overline{A_0\overline{(-\zeta_1)}^{2s}+iA_1\overline{(-\zeta_1)}^{2s-1}\overline{\zeta_2}+A_2\overline{(-\zeta_1)}^{2s-2}\overline{\zeta_2}^{2}+\dots +A_{2s}\overline{(\zeta_2)}^{2s}}\nonumber\\
	=\psi^{(I)}_{2s}(\zeta_1, \zeta_2).
\end{eqnarray}
Similarly, $\mathcal{C}_2^{(2)}\psi^{(I)}_{2s}(\zeta_1, \zeta_2)=\psi^{(I)}_{2s}(\zeta_1, \zeta_2)$.

\vspace{0.5cm}

\textbf{Case-II} :  $\vert m\vert\in\{2s-1 : s\in {Z^+}\}$
Similar argument as given above can lead to the following eigenfunction for odd $m$
\begin{eqnarray}
	\psi^{(II)}_{2s-1}(\zeta_1, \zeta_2)=B_0\zeta_1^{2s-1}+iB_1\zeta_1^{2s-2}\zeta_2+B_2\zeta_1^{2s-3}\zeta_2^{3}+\dots +iB_{2s-1}\zeta_2^{2s-1}
\end{eqnarray} 
where, $\{B_l : l=0\dots 2s-1\}$ are real functions of some real eigenvalue $\lambda_1$ with $B_0=1$.

Now actions of $\{\mathcal{C}_2^{(j)} : j=1, 2\}$ on $\psi^{(II)}_{2s-1}(\zeta_1, \zeta_2)$ are given by
\begin{eqnarray}
	\mathcal{C}_2^{(1)}\psi^{(II)}_{2s-1}(\zeta_1, \zeta_2)=-\psi^{(II)}_{2s-1}(\zeta_1, \zeta_2)\:\:{\rm and}\:\:\mathcal{C}_2^{(2)}\psi^{(II)}_{2s-1}(\zeta_1, \zeta_2)=\psi^{(II)}_{2s-1}(\zeta_1, \zeta_2).
\end{eqnarray}
{\it \bf{Remark-I}} : 
	It is observed that the eigen functions for even $m$ are $\partial_{\mathcal{PT}}$ symmetric in either of the variables whereas, for odd $m$ the eigenfunctions are symmetric in one variable $(\zeta_2)$ and anti-symmetric in the other $(\zeta_1)$. It can also be shown that $\mathcal{C}_2\psi^{(I)}_{2s}(\zeta_1, \zeta_2)\neq\psi^{(I)}_{2s}(\zeta_1, \zeta_2)$ and $\mathcal{C}_2\psi^{(II)}_{2s-1}(\zeta_1, \zeta_2)\neq\psi^{(II)}_{2s-1}(\zeta_1, \zeta_2)$ implying the fact that neither of them has any global $\mathcal{PT}$ symmetry.
	
{\bf{Remark-II}} : 
	If the eigenvalue has any nonzero complex part the antilinear action of the operators representing $\partial_{\mathcal{PT}}$ symmetry replaces the coefficients $\{A_l\}$ or $\{B_l\}$ by their respective complex conjugates thus destroying the symmetry of the eigenfunctions. This phenomenon may termed as \textbf{$\partial_{\mathcal{PT}}$ symmetry breaking}. 
	
	 It is obvious that the eigenvalues have a strong parametric dependence on the values of the parameters $\gamma$ and $\alpha$ an issue that may be discussed elsewhere. Considering $m=2$ and $\gamma=\alpha=\frac{1}{2}$ the eigenvalues are $\lambda_1=3.87513, \lambda_{2, 3}=0.06244\pm 0.71569 i$. This means, according to the above discussion, the states corresponding to $\lambda_1$ is a  $\partial_{\mathcal{PT}}$ symmetric state but those corresponding to $\lambda_2$ and $\lambda_3$ are symmetry breaking states. On the other hand, for $\alpha=4=8\gamma$ the eigenvalues are $(0.06375, 13.96387, 17.97237)$ and consequently, all the states are {$\partial_{\mathcal{PT}}$ symmetric.
			
			In the above discussion only the cases with non-degenerate eigenvalues have been considered. The situation with degenerate eigenvalues may be an interesting area of investigation. 
		 
\section{Conclusion}
The present investigation deals with a new kind of symmetry operator that acts on operators or functions in a variable specific way. In this article, the presence of such symmetry known as Partial $\mathcal{PT}$ symmetry is understood in a typical Fock space setting for a two-boson Bose-Hubbard type Hamiltonian with one purely imaginary interaction term. The symmetry operators are represented as \textbf{Weighted Composition Conjugations} acting on a \textbf{Reproducing Kernel Hilbert Space}. The existence of such symmetry as well as the possibility of breaking of such symmetry are found to have direct correspondence with the reality of eigenvalues of the Hamiltonian.

\end{document}